\documentclass{article}
\usepackage{amssymb}
\usepackage{amsmath}
\usepackage[T1]{fontenc}
\usepackage[T1]{fontenc}
\usepackage[T1]{fontenc}

\input{tcilatex}
\begin{document}

\title{Electromagnetic Induced Gravitational Perturbations }
\author{T.M. Adamo$^{1}$, E.T. Newman$^{2}$ \\
$^{1}$Dept of Mathematics, University of Pittsburgh, Pgh., PA 15260\\
$^{2}$Dept of Physics and Astronomy, University of Pittsburgh, Pgh., PA \ \
\ \ \ \ \\
\ \ 15260}
\date{July 18, 2008}
\maketitle

\begin{abstract}
We study the physical consequences of two diffferent but closely related
perturbation schemes applied to the Einstein-Maxwell equations. In one case
the starting space-time is flat while in the other case it is Schwarzschild.
In both cases the perturbation is due to a combined electric and magnetic
dipole field. \ We can see, within the Einstein-Maxwell equations a variety
of physical consequences. They range from induced gravitational
energy-momentum loss, to a well defined spin angular momentum with its loss
and a center-of-mass with its equations of motion.
\end{abstract}

\section{Introduction}

Recently, using the spin-coefficient (SC) formalism\cite{NPF}, a
perturbation scheme for a simple model using the Einstein-Maxwell equations
was described\cite{AdamoNewman,Bramson}. \ Pure electric and magnetic dipole
radiation was considered as a first order perturbation off a flat space
background. \ It was shown that this in fact leads to a second-order
perturbation in the gravitational field (the Weyl tensor), and in turn a
perturbed second order metric. \ Some interesting results were obtained by
simply looking at the asymptotic Weyl tensor; these included the existence
of a Bondi news function created by the dipole radiation with the
accompanying classical Bondi gravitational and electromagnetic energy-loss.
A pretty result was that one could identify, in the Bianchi Identities, the
classical electromagnetic angular momentum loss\cite{LL}.

In the present work, we apply this scheme to more complicated perturbations.
\ We find, perturbatively, two different versions of what could be loosely
characterized as generalized Reissner-Nordstr\"{o}m space-times; that is,
metrics with a mass and Coulomb charge, but now with electromagnetic dipole
radiation. \ In one case (perturbations off Schwarzschild) we consider the
mass to be zeroth order and the charge and dipole fields to be first order.
In the second case (perturbations off flat space) we consider the mass, the
charge and dipole field all to be first order.

In Section II, the Maxwell equations, for both perturbation types, are first
integrated. (It does not actually matter in which background [flat or
Schwarzschild[ this integration occurs for these two cases.) \ Next,
(Sections III and IV) with the Maxwell field in the stress tensor as the
source,\ we integrate the Bianchi identities obtaining the radial and
non-radial behavior of the Weyl tensor. We then probe further the asymptotic
behavior of the Weyl tensor, in particular looking at the angular momentum
loss and Bondi energy-momentum loss theorem as well as dynamical equations
for the motion of the center of mass and charge. Different physical
consequences for the two different perturbations are discovered.

For completeness, in Section V the full behavior of the second order metric
is presented. Sec. VI contains the discussion, while an appendix contains
full expressions for the Weyl tensor and spin coefficients.

\section{The Maxwell Field}

We work in the Bondi coordinate system $(u,r,\zeta ,\bar{\zeta})$ , where $u$
labels the light cones,$\mathfrak{C,}$ with apex on a time-like world-line, $%
r$ is the affine parameter along the null geodesics, and $\zeta =\cot
(\theta /2)e^{i\phi }$ is the complex stereographic angle labeling the null
geodesics on $\mathfrak{C}$. \ Furthermore, we choose the Bondi null tetrad $%
\{l^{a},n^{a},m^{a},\bar{m}^{a}\}$ such that the vector $l^{a}$ is tangent
to the null geodesic congruences. \ This, along with the choice of $l^{a}$
as a gradient, fixes the spin coefficients $\kappa ,\pi ,\varepsilon ,\rho $%
, and $\tau $ as \cite{NewmanTod}: $\kappa =\pi =\varepsilon =0$, $\rho =%
\bar{\rho}$, and $\tau =\bar{\alpha}+\beta $. \ In the Schwarzschild
space-time, this fixes the set of SCs as\cite{Schwarzschild}: 
\begin{eqnarray}
\kappa &=&\pi =\varepsilon =\sigma =\tau =\nu =\lambda =0,  \label{1} \\
\rho &=&-r^{-1},  \notag \\
\alpha &=&-\frac{\zeta }{2r}\equiv \frac{\alpha ^{0}}{r},  \notag \\
\beta &=&\frac{\bar{\zeta}}{2r}\equiv \frac{\beta ^{0}}{r},  \notag \\
\gamma &=&\frac{\sqrt{2}G}{r^{2}c^{2}}M_{S}  \notag \\
\mu &=&-\frac{1}{r}+\frac{2\sqrt{2}G}{r^{2}c^{2}}M_{S}.  \notag
\end{eqnarray}

The flat space Minkowski set of SCs are obtained simply by setting the
Schwarzschild mass, $M_{S}$, equal to zero. \ In both cases it can be seen
that the radial and non-radial Maxwell equations are the same\cite%
{NewmanTod,NewmanEM}, allowing us to write down the desired solution which
contains both a Coulomb charge and radiating electromagnetic dipole given
by: 
\begin{eqnarray}
\phi _{0} &=&\frac{\phi _{0}^{0}}{r^{3}}\equiv \frac{2D^{i}}{r^{3}}Y_{1i}^{1}
\label{2} \\
\phi _{1} &=&\frac{\phi _{1}^{0}}{r^{2}}+\frac{\phi _{1}^{1}}{r^{3}}\equiv 
\frac{q}{r^{2}}+\frac{\sqrt{2}D^{i\prime }}{r^{2}}Y_{1i}^{0}-\frac{D^{i}}{
r^{3}}Y_{1i}^{0}  \notag \\
\phi _{2} &=&\frac{\phi _{2}^{0}}{r}+\frac{\phi _{2}^{1}}{r^{2}}+\frac{\phi
_{2}^{2}}{r^{3}}\equiv -\frac{2D^{i\prime \prime }}{r}Y_{1i}^{-1}+\frac{2%
\sqrt{2}D^{i\prime }}{r^{2}}Y_{1i}^{-1}-\frac{D^{i}}{2r^{3}}Y_{1i}^{-1}. 
\notag
\end{eqnarray}
\ 

The three-vector $D^{i}$ is our complex dipole moment and can be written as
the complex superposition of its real electric and magnetic

\noindent parts:

\begin{equation}
D^{i}=D_{E}^{i}+iD_{M}^{i},  \label{3}
\end{equation}
while $q$ is our Coulomb charge. \ At this point, $D^{i}$ may be regarded as
an arbitrary function of the retarded time, $u_{r}\equiv \sqrt{2}u$. \ In
both perturbation calculations that follow, we will treat both $D^{i}$ and $%
q $ as being first-order quantities in the perturbation. \ All calculations
will be performed to second order. In addition we only keep the $l=0,1,2$
spherical harmonics.

\textbf{Remark }Other papers \cite{RobinsonTrautman,UCF,PhysicalContent}have
often written $D^{i}=q\xi ^{i}$, where $\xi ^{i}$ is a complex position
vector. \ Due to our perturbation formalism, we can make no such
identification at this point in the calculation.

\textbf{Remark }Throughout this paper, we denote differentiation with
respect to Bondi time as: $\partial _{u}()=(^{\cdot })$, while
differentiation with respect to the retarded time, $u_{r}=\sqrt{2}u,$ is
given as $\partial _{u_{r}}()=(^{\prime })$. \ Thus, we see that $(^{\cdot
})=\sqrt{2}(^{\prime })$. \ In later sections, when we want to restore units
where $c\neq 1$, we must take $(^{\prime })\rightarrow c^{-1}(^{\prime })$.
The gravitational coupling constant is $k=2Gc^{-4}.$

\section{Schwarzschild Background Perturbation}

We now consider the Maxwell field given by Eq.(\ref{2}) to be a first order
perturbation in the background of the Schwarzschild space-time with the
spin-coefficients given by Eq.(\ref{1}). \ In what follows, we carry this
information, via the stress tensor, into the Weyl tensor by integrating the
SC form of the Bianchi identities. Looking at the asymptotic Weyl tensor
components we study the physical consequences (i.e., mass, momentum, angular
momentum and equations of motion seen at null infinity) of our model.

\subsection{The Radial and Non-radial Bianchi Identities}

We are seeking a solution of the Bianchi Identities which is driven \textit{%
exclusively} by the original Schwarzschild mass (zeroth order) and
electromagnetic perturbation; i.e. with no gravitational degrees of freedom.
This leads to $\psi _{0}=0$. (This step has been taken by others, e.g., \cite%
{AdamoNewman,Bramson}). \ The radial Bianchi identities are then given by 
\cite{NewmanTod}: 
\begin{equation}
\frac{\partial \psi _{1}}{\partial r}+\frac{4\psi _{1}}{r}=\frac{\text{ \={%
\dh }}\psi _{0}}{r}+\frac{5k\phi _{0}^{0}\bar{\phi}_{1}^{0}}{r^{6}}+\frac{k%
\text{\dh }(\phi _{0}^{0}\bar{\phi}_{0}^{0})-4k\phi _{0}^{0}\text{\dh }\bar{%
\phi}_{0}^{0}}{r^{7}},  \label{4}
\end{equation}

\begin{eqnarray*}
\frac{\partial \psi _{2}}{\partial r} &=&-\frac{3\psi _{2}}{r}+\frac{\text{%
\={\dh }}\psi _{1}}{r}-\frac{2k\phi _{1}^{0}\bar{\phi}_{1}^{0}}{r^{5}}+\frac{%
4k}{3r^{6}}{\large [}\phi _{1}^{0}\text{\dh }\bar{\phi}_{0}^{0}+\bar{\phi}%
_{1}^{0}\text{\={\dh }}\phi _{0}^{0}-\frac{1}{2}\text{\dh }(\phi _{1}^{0}%
\bar{\phi}_{0}^{0})+\frac{1}{4}\text{\={\dh }}(\phi _{0}^{0}\bar{\phi}%
_{1}^{0}){\large ]} \\
&&-\frac{k}{3r^{7}}{\large [}\frac{5}{2}\text{\={\dh }}\phi _{0}^{0}\text{%
\dh }\bar{\phi}_{0}^{0}-\text{\dh }(\bar{\phi}_{0}^{0}\text{\={\dh }}\phi
_{0}^{0})+\frac{1}{2}\text{\={\dh }}(\phi _{0}^{0}\text{\dh }\bar{\phi}%
_{0}^{0})+\phi _{0}^{0}\bar{\phi}_{0}^{0}{\large ]}-k\Delta \left( \frac{%
\phi _{0}^{0}\bar{\phi}_{0}^{0}}{3r^{6}}\right) \\
&&+\frac{2\sqrt{2}kG}{c^{2}}\frac{M_{S}\phi _{0}^{0}\bar{\phi}_{0}^{0}}{r^{8}%
},
\end{eqnarray*}

\begin{eqnarray*}
\frac{\partial \psi _{3}}{\partial r} &=&-\frac{2\psi _{3}}{r}+\frac{\text{%
\={\dh }}\psi _{2}}{r}{\small -}\frac{k\phi _{2}^{0}\bar{\phi}_{1}^{0}}{r^{4}%
}+\frac{k}{3r^{5}}\left[ 2\phi _{2}^{0}\text{\dh }\bar{\phi}_{0}^{0}+4\bar{%
\phi}_{1}^{0}\text{\={\dh }}\phi _{1}^{0}{\small -}\text{\dh }(\phi _{2}^{0}%
\bar{\phi}_{0}^{0})+2\text{\={\dh }}(\phi _{1}^{0}\bar{\phi}_{1}^{0})\right]
\\
&&-\frac{k}{3r^{6}}\left[ \frac{5}{2}\text{\={\dh }}\phi _{1}^{0}\text{\dh }%
\bar{\phi}_{0}^{0}+\frac{5}{4}\bar{\phi}_{1}^{0}\text{\={\dh }}^{2}\phi
_{0}^{0}-\text{\dh }(\bar{\phi}_{0}^{0}\text{\={\dh }}\phi _{1}^{0})+\text{ 
\={\dh }}(\phi _{1}^{0}\text{\dh }\bar{\phi}_{0}^{0})+\text{\={\dh }}(\bar{%
\phi}_{1}^{0}\text{\={\dh }}\phi _{0}^{0})\right] \\
&&+\frac{k}{12r^{7}}\left[ 2\text{\={\dh }}(\text{\={\dh }}\phi _{0}^{0}%
\text{\dh }\bar{\phi}_{0}^{0})-\text{\dh }(\bar{\phi}_{0}^{0}\text{\={\dh }}%
^{2}\phi _{0}^{0})+3\text{\dh }\bar{\phi}_{0}^{0}\text{\={\dh }}^{2}\phi
_{0}^{0}\right] -\frac{2k}{3}\Delta \left( \frac{\phi _{1}^{0}\bar{\phi}%
_{0}^{0}}{r^{5}}-\frac{\bar{\phi}_{0}^{0}\text{\={\dh }}\phi _{0}^{0}}{2r^{6}%
}\right) \\
&&+\frac{4\sqrt{2}kG}{3c^{2}}M_{S}\left( \frac{\phi _{1}^{0}\bar{\phi}%
_{0}^{0}}{r^{7}}-\frac{\bar{\phi}_{0}^{0}\text{\={\dh }}\phi _{0}^{0}}{2r^{8}%
}\right) ,
\end{eqnarray*}

\begin{eqnarray*}
\frac{\partial \psi _{4}}{\partial r} &=&-\frac{\psi _{4}}{r}+\frac{\text{ 
\={\dh }}\psi _{3}}{r}+\frac{\text{\={\dh }}(\phi _{2}^{0}\bar{\phi}_{1}^{0})%
}{r^{4}}-\frac{k}{2r^{5}}\left[ \text{\={\dh }}(\phi _{2}^{0}\text{\dh }\bar{%
\phi}_{0}^{0})+2\text{\={\dh }}(\bar{\phi}_{1}^{0}\text{\={\dh }}\phi
_{1}^{0})-2\phi _{2}^{0}\bar{\phi}_{0}^{0}\right] \\
&&+\frac{k}{4r^{6}}\left[ 2\text{\={\dh }}(\text{\={\dh }}\phi _{1}^{0}\text{
\dh }\bar{\phi}_{0}^{0})+\text{\={\dh }}(\bar{\phi}_{1}^{0}\text{\={\dh }}%
^{2}\phi _{0}^{0})-4\bar{\phi}_{0}^{0}\text{\={\dh }}\phi _{1}^{0}\right] -%
\frac{k\left[ \text{\={\dh }}(\text{\dh }\bar{\phi}_{0}^{0}\text{\={\dh }}%
^{2}\phi _{0}^{0})-2\bar{\phi}_{0}^{0}\text{\={\dh }}^{2}\phi _{0}^{0}\right]
}{8r^{7}} \\
&&-k\Delta \left( \frac{\phi _{2}^{0}\bar{\phi}_{0}^{0}}{r^{4}}-\frac{\bar{%
\phi}_{0}^{0}\text{\={\dh }}\phi _{1}^{0}}{r^{5}}+\frac{\bar{\phi}_{0}^{0}%
\text{\={\dh }}^{2}\phi _{0}^{0}}{4r^{6}}\right) -\frac{2\sqrt{2}kG}{c^{2}}%
M_{S}\left( \frac{\phi _{2}^{0}\bar{\phi}_{0}^{0}}{r^{6}}-\frac{\bar{\phi}%
\text{\={\dh }}\phi _{1}^{0}}{r^{7}}+\frac{\bar{\phi}_{0}^{0}\text{\={\dh }}%
^{2}\phi _{0}^{0}}{4r^{8}}\right) .
\end{eqnarray*}

The differential operator $\Delta $ is given by: $\Delta \equiv \partial
_{u}-\partial _{r}+r^{-2}c^{-2}2\sqrt{2}GM_{S}\partial _{r}$.

Integration produces the following results. (We relegate the complicated
terms, $\mathcal{A}_{i},$ involving higher $r$-dependence to the appendix,
as these are largely unnecessary for the calculations of particular interest
in this paper): 
\begin{equation}
\psi _{1}=\frac{\psi _{1}^{0}}{r^{4}}+\mathcal{A}_{1}  \label{8}
\end{equation}

\begin{equation}
\psi _{2}=\frac{\psi _{2}^{0}}{r^{3}}+\mathcal{A}_{2}  \label{9}
\end{equation}

\begin{equation}
\psi _{3}=\frac{\psi _{3}^{0}}{r^{2}}+\mathcal{A}_{3}  \label{10}
\end{equation}

\begin{equation}
\psi _{4}=\frac{\psi _{4}^{0}}{r}+\mathcal{A}_{4.}  \label{11}
\end{equation}

Here, the $\psi _{1}^{0},\psi _{2}^{0},\psi _{3}^{0}$ and $\psi _{4}^{0}$
are $r~$- independent functions of integration, which can be determined from
the non-radial Bianchi identities, which take the form \cite{NewmanTod} : 
\begin{equation}
\text{\dh }\psi _{1}^{0}=3k\phi _{0}^{0}\bar{\phi}_{2}^{0},  \label{12}
\end{equation}

\begin{equation}
\dot{\psi}_{1}^{0}=-\text{\dh }\psi _{2}^{0}+2k\phi _{1}^{0}\bar{\phi}
_{2}^{0},  \label{13}
\end{equation}

\begin{equation}
\dot{\psi}_{2}^{0}=-\text{\dh }\psi _{3}^{0}+k\phi _{2}^{0}\bar{\phi}
_{2}^{0},  \label{14}
\end{equation}

\begin{equation}
\dot{\psi}_{3}^{0}=-\text{\dh }\psi _{4}^{0}.  \label{15}
\end{equation}

The integration of Eqs.(\ref{12})-(\ref{15}) is relatively straight forward,
basically entailing the comparison of spherical harmonic coefficients. \ We
make considerable use of Clebsh-Gordon expansions \cite{Harmonics} of the
quadratic terms in the Maxwell field.\ This process applied to Eq.(\ref{12})
yields: 
\begin{equation}
\psi _{1}^{0}=3kD^{i}\bar{D}^{j\prime \prime }Y_{2ij}^{1}+\psi
_{1}^{0\,k}Y_{1k}^{1}.  \label{16}
\end{equation}

The vector $\psi _{1}^{0\,k}(u_{r})$ emerges as an unknown integration
factor in much the same way that $\psi _{1}^{0}$ emerged from the
integration of the radial equations. \ Shortly, however, we will be able to
determine $\psi _{1}^{0\,k}$ up to constants. \ From Eq.(\ref{13}), we find: 
\begin{eqnarray}
\psi _{2}^{0} &=&-\frac{2\sqrt{2}G}{c^{2}}M_{S}+\Upsilon _{\epsilon }+%
{\large (}\frac{\sqrt{2}}{2}\psi _{1}^{0\,k\prime }+2ki\bar{D}^{i\prime
\prime }D^{j\prime }\epsilon _{ijk}{\large )}Y_{1k}^{0}  \label{17} \\
&&+2kq\bar{D}^{k\prime \prime }Y_{1k}^{0}+\sqrt{2}k{\large (}\frac{(D^{i}%
\bar{D}^{j\prime \prime })^{\prime }}{2}+\frac{\bar{D}^{i\prime \prime
}D^{j\prime }}{3}{\large )}Y_{2ij}^{0},  \notag
\end{eqnarray}%
where the $l=0$ function $\Upsilon _{\epsilon }(u_{r})$ is a (first-order)
function of integration. \ Considering Eq.(\ref{14}), we recognize that as $%
\psi _{3}^{0}$ is a spin weighted $s=-1$ quantity it has no $l=0$ harmonic
contribution. This allows us to both obtain $\psi _{3}^{0}$ as well as place
a restriction on $\Upsilon _{\epsilon }$: 
\begin{eqnarray}
\psi _{3}^{0} &=&\left( \sqrt{2}ki\bar{D}^{i\prime \prime }D^{j\prime \prime
}\epsilon _{ijk}+2\sqrt{2}ki(D^{i\prime }\bar{D}^{j\prime \prime })^{\prime
}\epsilon _{ijk}\,-\psi _{1}^{0\,k\prime \prime }\right) Y_{1k}^{-1}
\label{18} \\
&&-2\sqrt{2}kq\bar{D}^{k\prime \prime \prime }Y_{1k}^{-1}+k\left( \frac{1}{3}%
\bar{D}^{i\prime \prime }D^{j\prime \prime }-(D^{i}\bar{D}^{j\prime \prime
})^{\prime \prime }-\frac{2}{3}(D^{i\prime }\bar{D}^{j\prime \prime
})^{\prime }\right) Y_{2ij}^{-1},  \notag
\end{eqnarray}

\begin{equation}
\dot{\Upsilon}_{\epsilon }=\sqrt{2}\Upsilon _{\epsilon }{}^{\prime }=\frac{4k%
}{3}D^{i\prime \prime }\bar{D}^{j\prime \prime }\delta _{ij}.  \label{19}
\end{equation}

Finally, we determine both $\psi _{1}^{0\,k}$ and $\psi _{4}^{0}$ from Eq.( %
\ref{15}). We recall that $\psi _{4}^{0}$ is an $s=-2$ quantity and hence
does not contain an $l=1$ harmonic. The $l=1$ harmonic contributions to the
equation thus must vanish yielding a differential condition on $\psi
_{1}^{0\,k}(u_{r})$; the remaining part yields the determination of $\psi
_{4}^{0}$: 
\begin{equation}
\psi _{4}^{0}=\sqrt{2}k\left[ (D^{i}\bar{D}^{j\prime \prime })^{\prime
\prime \prime }+\frac{2}{3}(D^{i\prime }\bar{D}^{j\prime \prime })^{\prime
\prime }-\frac{1}{3}(\bar{D}^{i\prime \prime }D^{j\prime \prime })^{\prime }%
\right] Y_{2ij}^{-2},  \label{20}
\end{equation}

\begin{equation}
\psi _{1}^{0\,k\prime \prime \prime }=\sqrt{2}ki[(\bar{D}^{i\prime \prime
}D^{j\prime \prime })^{\prime }\epsilon _{ijk}+2(D^{i\prime }\bar{D}%
^{j\prime \prime })^{\prime \prime }\epsilon _{ijk}\,+2iq\bar{D}^{k\prime
\prime \prime \prime }].  \label{20a}
\end{equation}

For later use we decompose Eq.(\ref{20a}) into its real and imaginary parts
using $\psi _{1}^{0\,k}=\psi _{1R}^{0\,k}+i\psi _{1I}^{0\,k}:$ 
\begin{eqnarray}
\psi _{1}^{0\,}{}_{R}^{k\prime \prime \prime } &=&2\sqrt{2}k(D_{M}^{j\prime
}D_{E}^{i\prime })^{\prime \prime \prime }\epsilon _{ijk}-\sqrt{2}%
k(D_{E}^{i\prime \prime }D_{M}^{j\prime \prime })^{\prime }\epsilon _{ijk}-2%
\sqrt{2}kqD_{E}^{k\prime \prime \prime \prime }  \label{20b} \\
\psi _{1I}^{0\,k\prime \prime \prime } &=&2\sqrt{2}k[(D_{E}^{i\prime
}D_{E}^{j\prime \prime }+D_{M}^{i\prime }D_{M}^{j\prime \prime })\epsilon
_{ijk}+qD_{M}^{k\prime \prime }]^{\prime \prime }.  \label{20c}
\end{eqnarray}

Note that the latter equation can be immediately integrated twice as

\begin{equation}
\psi _{1I}^{0\,k\prime }=2\sqrt{2}k[(D_{E}^{i\prime }D_{E}^{j\prime \prime
}+D_{M}^{i\prime }D_{M}^{j\prime \prime })\epsilon _{ijk}+qD_{M}^{k\prime
\prime }],  \label{20d}
\end{equation}
where we have taken the two constants of integration to vanish while the
first can be integrated once as:

\begin{equation}
\psi _{1}^{0\,}{}_{R}^{k\prime \prime }=\sqrt{2}k[2(D_{M}^{j\prime
}D_{E}^{i\prime })^{\prime \prime }-D_{E}^{i\prime \prime }D_{M}^{j\prime
\prime }]\epsilon _{ijk}-2\sqrt{2}kqD_{E}^{k\prime \prime \prime }.
\label{20e}
\end{equation}

\textbf{Remark: } Bramson \cite{Bramson},improperly set the $\psi
_{1}^{0\,k}=0,$ apparently overlooking this \noindent differential condition.

\textbf{Remark: }There is a further equation (a reality condition) on $\psi
_{2}^{0}$ that plays a very important role. It however involves a further
variable (the spin-coefficient $\sigma ^{0}$, \ i.e., \ the Bondi shear) for
its description.

\subsection{Reality Conditions}

To more easily see the physical content in our equations, we now introduce
the $c$ in all time derivatives, i.e., $(^{\prime })\rightarrow
c^{-1}(^{\prime })$.

We turn to the Bondi mass aspect where its reality forces further
restrictions on our variables. \ To proceed, we must first know the shear
i.e., the spin coefficient $\sigma $. \ A more detailed discussion of the
spin coefficients is given in Section V and the Appendix, but it can be seen 
\cite{AdamoNewman} that the behavior of $\sigma $ does not change with the
switch from flat-space to a Schwarzschild background or the addition of a
Coulomb charge. \ Thus, we find\cite{AdamoNewman}: 
\begin{eqnarray}
\sigma &=&\frac{\sigma ^{0}}{r^{2}},  \label{22} \\
\psi _{4}^{0} &=&-2c^{-2}\bar{\sigma}^{0\prime \prime }  \label{22a}
\end{eqnarray}

\begin{equation}
\sigma ^{0}=\frac{\sqrt{2}k}{2c^{3}}\left[ \frac{1}{3}\dint (D^{i\prime
\prime }\bar{D}^{j\prime \prime })du_{r}-(\bar{D}^{i}D^{j\prime \prime
})^{\prime }-\frac{2}{3}(\bar{D}^{i\prime }D^{j\prime \prime })\right]
Y_{2ij}^{2}.  \label{23}
\end{equation}

Using Eqs.(\ref{22}) and (\ref{23}), the Bondi mass aspect\cite%
{Bondi,NewmanTod} is given by 
\begin{equation}
\Psi =\psi _{2}^{0}+\text{\dh }^{2}\bar{\sigma}^{0}+\sqrt{2}c^{-1}\sigma
^{0}(\bar{\sigma}^{0})^{\prime },  \label{24}
\end{equation}%
and is subject to the reality condition: 
\begin{equation}
\Psi =\bar{\Psi}.  \label{25}
\end{equation}

As can be seen from Eq.(\ref{23}), the quantity $\sigma ^{0}$ is second
order in the perturbation, so that it can be neglected in the definition of
the mass aspect in Eq.(\ref{24}). \ If we expand the mass aspect in
spherical harmonics, 
\begin{equation*}
\Psi =\Psi ^{0}+\Psi ^{i}Y_{1i}^{0}+\Psi ^{ij}Y_{2ij}^{0}+\cdots ,
\end{equation*}%
then from Bondi\cite{Bondi,NewmanTod}, the $l=0$ and $l=1$ terms, i.e., 
\begin{equation}
\Psi ^{0}=-\frac{2\sqrt{2}G}{c^{2}}M_{S}+\Upsilon _{\epsilon }=-\frac{2\sqrt{%
2}G}{c^{2}}M_{B}  \label{26}
\end{equation}

\begin{equation}
\Psi ^{i}=-\frac{6G}{c^{3}}P^{i}  \label{27}
\end{equation}
are, up to numerical factors, interpreted as the Bondi mass and linear
three-momentum respectively.

From Eq.(\ref{24}) using Eqs.(\ref{17}) and (\ref{23}), we have that: 
\begin{eqnarray}
\Psi &=&-\frac{2\sqrt{2}G}{c^{2}}M_{S}+\Upsilon _{\epsilon }+\left( \frac{%
\sqrt{2}}{2c}\psi _{1}^{0\,k\prime }+\frac{2ki}{c^{3}}\bar{D}^{i\prime
\prime }D^{j\prime }\epsilon _{ijk}\right) Y_{1k}^{0}  \label{28} \\
&&+\frac{2kq}{c^{2}}\bar{D}^{k\prime \prime }Y_{1k}^{0}+\frac{\sqrt{2}k}{%
6c^{3}}\dint (D^{i\prime \prime }\bar{D}^{j\prime \prime })du_{r}Y_{2ij}^{0}.
\notag
\end{eqnarray}%
By the symmetry on the $l=2$ contribution, it follows that $\Psi ^{ij}$ is
real. \ From the $l=0$ contribution, we see that the Bondi mass is real: 
\begin{equation}
-\frac{2\sqrt{2}G}{c^{2}}M_{S}+\Upsilon _{\epsilon }=-\frac{2\sqrt{2}G}{c^{2}%
}M_{S}+\bar{\Upsilon}_{\epsilon }=-\frac{2\sqrt{2}G}{c^{2}}M_{B}.  \label{29}
\end{equation}%
The reality condition on the $l=1$ condition is a bit more complicated. \
First, we decompose the dipole $D^{i}$ into its electric and magnetic parts, 
\begin{equation*}
D^{i}=D_{E}^{i}+iD_{M}^{i},
\end{equation*}%
and write: $\psi _{1}^{0\,k}$\ \ $=\psi _{1R}^{0\,k}$\ \ $+i\psi
_{1I}^{0\,k} $\ \ . \ Then the condition for $\Psi ^{i}=\bar{\Psi}^{i}$
yields two relations; one on the real and one on the imaginary part of $\psi
_{1}^{0\,k} $. \ We find for the vanishing of the imaginary part $\psi
_{1}^{0\,k},$ 
\begin{equation}
\psi _{1I}^{0\,k\prime }=2\sqrt{2}kc^{-1}qD_{M}^{k\prime \prime }+2\sqrt{2}%
kc^{-2}(D_{E}^{i\prime }D_{E}^{j\prime \prime }+D_{M}^{i\prime
}D_{M}^{j\prime \prime })\epsilon _{ijk},  \label{30}
\end{equation}%
which is identical to the earlier derived, Eq.(\ref{20d}). The real part
leads to an expression for the Bondi linear momentum

\begin{equation}
\Psi ^{k}=-\frac{6G}{c^{3}}P^{k}=\left[ \frac{\sqrt{2}}{2c}\psi
_{1R}^{0\,k}+\,\frac{2kq}{c^{2}}D_{E}^{k\prime }-\frac{2k}{c^{3}}%
(D_{M}^{j\prime }D_{E}^{i\prime })\epsilon _{ijk}\right] ^{\prime },
\label{31}
\end{equation}
or

\begin{equation}
P^{k}=-\left[ \frac{c^{2}\sqrt{2}}{12G}\psi _{1R}^{0\,k\prime }+\,\frac{2q}{%
3c^{3}}D_{E}^{k\prime \prime }-\frac{2}{3c^{4}}(D_{M}^{j\prime
}D_{E}^{i\prime })^{\prime }\epsilon _{ijk}\right] .  \label{31a}
\end{equation}%
By taking the $u_{r}$ derivative and simplifying via Eq.(\ref{20e}), we
obtain the electromagnetic momentum flux law,

\begin{equation}
P^{k\prime }=\frac{1}{3}c^{-4}D_{E}^{i\prime \prime }D_{M}^{j\prime \prime
}\epsilon _{ijk}.  \label{31b}
\end{equation}

The imaginary part of the $l=1$ harmonic of $\psi _{1}^{0},$ i.e., $\psi
_{1I}^{0\,k},$ is often, \textit{in vacuum linear theory}, taken as
proportional to the total source angular momentum as viewed from infinity.
This becomes modified\cite{PhysicalContent} in the presence of a Maxwell
Field as 
\begin{equation}
\psi _{1I}^{0\,k}-2\sqrt{2}kqc^{-1}D_{M}^{k\prime }=-\frac{6\sqrt{2}G}{c^{3}}%
J^{k}.
\end{equation}%
Eq.(\ref{30}) is seen as the \textit{classical law of conservation of
angular momentum for electromagnetic dipole radiation}\cite{LL} or angular
momentum flux law: 
\begin{equation}
J^{k\prime }=\frac{2}{3c^{3}}(D_{E}^{i\prime \prime }D_{E}^{j\prime
}+D_{M}^{i\prime \prime }D_{M}^{j\prime })\epsilon _{ijk}.  \label{ang1}
\end{equation}

\subsection{Null Rotations and Equations of Motion}

We now construct a transformation (a \textit{null tetrad rotation around }$%
n^{a})$ to what we define as the \textit{complex center of mass\cite%
{PhysicalContent}}.

Though this is not the place to go into a detailed explanation \cite%
{Footprints,PhysicalContent,UCF} of the meaning of the term `\textit{complex
center of mass', }nor into the details how it can be calculated or found, a
brief explanation is in order.

In a given asymptotically flat space-time, the family of regular
asymptotically shear-free null geodesic congruences are determined by 1. the
Bondi asymptotic shear $\sigma ^{0}(u_{r},\zeta ,\bar{\zeta})$ and 2. an
arbitrary choice of a complex world-line that `lives' in the space of
complex Poincare transformations (complex Minkowski space), a subgroup of
the BMS group acting on $\mathfrak{I}^{+}.$ \{That an asymptotically shear
free null geodesic congruence picks out a complex world-line in complex
Minkowski space is the central fact in the present discussion. That it has
led to a series of remarkable results\cite{PhysicalContent,RadiationReaction}
is the defense\cite{3rdPrize} of its relevance.\} A \underline{\textit{%
particular} complex world-line \textit{can be chosen}} so that the
asymptotically defined center of mass and angular momentum both vanish on
it. The basic variable used to describe the asymptotically shear-free null
geodesic congruence is a (stereographic) angle field, $L(u_{r},\zeta ,\bar{%
\zeta})$ on $\mathfrak{I}^{+}$ that points backwards into the space-time
determining a past null direction at each point ($u_{r},\zeta ,\bar{\zeta}$)
of $\mathfrak{I}^{+}$. The angle field, for asymptotically shear-free
congruences, satisfies the differential equation \cite{Footprints} 
\begin{equation}
\text{\dh }L+LL,_{u}=\sigma ^{0}.  \label{L}
\end{equation}%
The solution to Eq.(\ref{L}), (accurate to our working order), is

\begin{equation}
L(u_{r},\zeta ,\bar{\zeta})=\xi ^{i}Y_{1i}^{1}-\frac{i}{2}\xi
^{i}v^{i}\epsilon _{ijk}Y_{1k}^{1}+\cdots ,  \label{LL}
\end{equation}
where $\xi ^{i}(u_{r})$ and $v^{i}=\xi ^{i\prime }$ are respectively the
arbitrary complex world-line and its velocity. Quadratic terms and high
harmonics \cite{PhysicalContent} have been omitted.

The transformation (null rotation) on $\mathfrak{I}^{+}$of the Bondi tetrad
( $l,n,m,\bar{m}$) to a new tetrad ($l^{*},n^{*},m^{*},\bar{m}^{*}$) (where
the $l^{*}$ is the null tangent vector to the asymptotically shear-free null
geodesic congruence) is given by

\begin{eqnarray}
l^{\ast } &=&l+\frac{L}{r}\bar{m}+\frac{\bar{L}}{r}m+\mathcal{O}(r^{-2})
\label{32} \\
m^{\ast } &=&m+\mathcal{O}(r^{-1})  \notag \\
n^{\ast } &=&n,  \notag
\end{eqnarray}

\textbf{Remark }We have gone from the Bondi tetrad frame with null vector $l$
that is twist free but has shear to a null vector $l^{\ast }$ that is
asymptotically shear-free but now possesses twist. The twist, $\Sigma ,$
given by 
\begin{equation*}
i\Sigma =\frac{1}{2}(\eth \bar{L}+\sqrt{2}L\bar{L}^{\prime }-\bar{\text{\dh }%
}L-\sqrt{2}\bar{L}L^{\prime }).
\end{equation*}%
It vanishes, in our approximation, when the world-line $\xi ^{i}$ is real.

This transformation induces a transformation of the asymptotic Weyl tensor
components. In particular the Weyl component $\psi _{1}^{0}$ transforms as

\begin{equation}
\psi _{1}^{0\ast }=\psi _{1}^{0}-3L\psi _{2}^{0}+\cdots ,  \label{psi_1}
\end{equation}

The \textit{basic physical idea} is that in linear theory the $l=1$ harmonic
component of $\psi _{1}^{0}$ is (usually) taken as proportional to the
complex center of mass, i.e., the (real) $center\,of$ $mass+\,i\,angular$ $%
momentum$. Our procedure is now to chose the arbitrary complex world-line so
that the $l=1$ harmonic component of $\psi _{1}^{0\ast }$ vanishes. \textit{%
The world-line so obtained is the complex center of mass. }We thus have to
solve

\begin{equation}
0=\psi _{1i}^{0}-3L\psi _{2}^{0}|_{i}+\cdots  \label{34}
\end{equation}
for $\xi ^{i}(u_{r}).$

Using Eqs.(\ref{16}), (\ref{17}), and (\ref{LL}), Eq.(\ref{34}) becomes: 
\begin{equation}
0=\psi _{1}^{0\,k}+\frac{6\sqrt{2}G}{c^{2}}M_{S}\xi ^{k}-\frac{3i\sqrt{2}G}{%
c^{2}}M_{S}\xi ^{i}v^{j}\epsilon _{ijk}.  \label{35}
\end{equation}
From Eqs. (\ref{20a}), (\ref{30}), and (\ref{31}), we saw that at least up
to initial conditions, the vector $\psi _{1}^{0\,k}$ is a second order
quantity in our perturbation framework. \ However by choosing \textit{%
time-independent first-order initial conditions}, $\psi _{1(0)}^{0k}$, we
see from Eq.(\ref{35}), that 
\begin{equation}
\xi ^{k}=-\frac{\sqrt{2}c^{2}\psi _{1(0)}^{0k}}{12GM_{S}},  \label{36}
\end{equation}
i.e., $\xi ^{k}$ is a constant vector and complex center of mass is at rest.
\ 

In this case, where $\xi ^{k}$ is a constant vector, the real part can be
set to zero by a Poincar\'{e} translation on $\mathfrak{I}^{+}$ so that 
\begin{equation}
\xi ^{k}=\xi _{I}^{k}=-\frac{c^{2}\psi _{1(0)}^{0k}}{6\sqrt{2}GM_{S}}.
\label{36a}
\end{equation}

Thus, we see that when the mass of the system is considered to be zeroth
order, the equation of motion for the center of mass is trivial: the center
of mass simply sits on the time axis. \ Physically, this can be thought of
in the following manner: if $M_{S}$ is the initial Schwarzschild mass, then
it is too \textquotedblleft heavy\textquotedblright\ for its motion to be
affected by the \textquotedblleft small\textquotedblright\ electromagnetic
perturbation given by Eq.(\ref{2}) (at least to second order in the
calculation). \ We see later that for small mass the situation is very
different.

\subsection{Physical Interpretations}

It turns out that there are a variety of physical interpretations \ - some
new and some old - to the results of this section.

First of all when the appropriate units are inserted, Eq.(\ref{19}) is
exactly the Bondi mass/energy loss equation

\begin{equation}
M_{B}^{\,\prime }=-\frac{2}{3c^{5}}(D_{E}^{i\,\prime \prime
}D_{E}^{i\,\prime \prime }+D_{M}^{i\,\prime \prime }D_{M}^{i\,\prime \prime
}).  \label{M'}
\end{equation}
At this approximation it coincides with the classical electromagnetic dipole
energy loss. At another approximation level there would be a forth-order
correction for gravitational energy loss via the square of the Bondi news
function, i.e., quadrupole radiation.

The Bondi momentum loss, given by Eq.(\ref{31b}), 
\begin{equation}
P^{k\prime }=\frac{1}{3c^{4}}D_{E}^{i\prime \prime }D_{M}^{j\prime \prime
}\epsilon _{ijk}  \label{P'}
\end{equation}
is just the electromagnetic momentum flux.

More interesting is Eq.(\ref{30}), (or (\ref{20d})): 
\begin{equation}
\psi _{1I}^{0\,k\prime }=2\sqrt{2}kqc^{-1}D_{M}^{k\prime \prime }+2\sqrt{2}%
kc^{-2}(D_{E}^{i\prime }D_{E}^{j\prime \prime }+D_{M}^{i\prime
}D_{M}^{j\prime \prime })\epsilon _{ijk}.  \label{30*}
\end{equation}

The imaginary part of the $l=1$ harmonic of $\psi _{1}^{0},$ i.e., $\psi
_{1I}^{0\,k},$ is often, \textit{in vacuum linear theory}, taken as
proportional to the total source angular momentum, $J^{k},$ as viewed from
infinity. This becomes modified\cite{PhysicalContent} in the presence of a
Maxwell Field as 
\begin{equation}
J^{k}=-\frac{c^{3}}{6\sqrt{2}G}\psi _{1I}^{0\,k}+\frac{2q}{3c^{2}}%
D_{M}^{k\prime }.  \label{J}
\end{equation}

Eq.(\ref{30*}) is seen as the \textit{classical law of conservation of
angular momentum\cite{LL} for electromagnetic dipole radiation}: 
\begin{equation}
J^{k\prime }=\frac{2}{3c^{3}}(D_{E}^{i\prime \prime }D_{E}^{j\prime
}+D_{M}^{i\prime \prime }D_{M}^{j\prime })\epsilon _{ijk}.  \label{J'}
\end{equation}

In the past, the imaginary part of the world-line vector, $\xi _{I}^{i},$
has been identified\cite{KerrNewman,PhysicalContent} with the intrinsic spin
associated with the asymptotic metric via the relationship 
\begin{equation}
S^{i}=M_{B}c\xi _{I}^{i}.  \label{S}
\end{equation}

We then see, from Eq.(\ref{36}), that the initial value of $\psi _{1}^{0k},$
i.e.,$\psi _{1(0)}^{0k}$ is proportional to the (constant) spin 
\begin{equation}
M_{S}c\xi _{I}^{k}=-\frac{c^{3}\psi _{1(0)}^{0k}}{6\sqrt{2}G}=S^{i}.
\label{Si}
\end{equation}%
so that the angular momentum, $J^{k},$ in Eq.(\ref{J}), consists of the sum
of three terms, the intrinsic spin, $S^{i},$ a complicated term involving
the integral over quadratic derivatives of the dipole moments and an unusual
term proportional to the derivative of the magnetic dipole moment, $\frac{2}{%
3}c^{-1}qD_{M}^{k\prime }$.

\section{Minkowski Background Perturbation}

We now consider an alternative perturbation scheme, namely perturbations off
Minkowski space-time. The gravitating mass now enters into the calculation
as a first order quantity in the mass term of the Weyl tensor component $%
\psi _{2}^{0}$. In the previous model the mass was large; now it will be
small.\ Most of the calculations of the previous section are essentially
identical in this framework, so we can proceed quickly toward the null
rotation calculations where new results do appear.

\subsection{Radial and Non-radial Bianchi Identities}

Once again, we set $\psi _{0}=0$ so that our solution is driven purely by
the perturbative quantities. \ The radial Bianchi identities are simply
those in flat Minkowski space-time. The results of their integration are the
same as those obtained earlier\cite{AdamoNewman} (but now with the addition
of a Coulomb charge) or by setting $M_{S}=0$ in Eqs.(\ref{4}). \ 

\ At the leading order in $r$, this yields radial behavior equivalent to
that found in the preceding section. The non-radial integrations give
virtually identical results: 
\begin{equation}
\psi _{1}^{0}=\psi _{1}^{0\,k}Y_{1k}^{1}+3kc^{-2}D^{i}\bar{D}^{j\prime
\prime }Y_{2ij}^{1},  \label{39}
\end{equation}

\begin{eqnarray}
\psi _{2}^{0} &=&\Upsilon _{\epsilon }+\left( \frac{2kq}{c^{2}}\bar{D}%
^{k\prime \prime }+\frac{\sqrt{2}}{2c}\psi _{1}^{0\,k\prime }+\frac{2ki}{%
c^{3}}\bar{D}^{i\prime \prime }D^{j\prime }\epsilon _{ijk}\right) Y_{1k}^{0}
\label{40} \\
&&+\frac{\sqrt{2}k}{c^{3}}\left( \frac{(D^{i}\bar{D}^{j\prime \prime
})^{\prime }}{2}+\frac{\bar{D}^{i\prime \prime }D^{j\prime }}{3}\right)
Y_{2ij}^{0},  \notag
\end{eqnarray}

\begin{eqnarray}
\psi _{3}^{0} &=&\left( \frac{\sqrt{2}ki}{c^{4}}\bar{D}^{i\prime \prime
}D^{j\prime \prime }\epsilon _{ijk}+\frac{2\sqrt{2}ki}{c^{4}}(D^{i\prime }%
\bar{D}^{j\prime \prime })^{\prime }\epsilon _{ijk}-c^{-2}\psi
_{1}^{0\,k\prime \prime }\right) Y_{1k}^{-1}  \label{41} \\
&&-\frac{2\sqrt{2}kq}{c^{2}}\bar{D}^{k\prime \prime
}Y_{1k}^{-1}+kc^{-4}\left( \frac{1}{3}\bar{D}^{i\prime \prime }D^{j\prime
\prime }-(D^{i}\bar{D}^{j\prime \prime })^{\prime \prime }-\frac{2}{3}%
(D^{i\prime }\bar{D}^{j\prime \prime })^{\prime }\right) Y_{2ij}^{-1}, 
\notag
\end{eqnarray}

\begin{equation}
\dot{\Upsilon}_{\epsilon }=\sqrt{2}\Upsilon _{\epsilon }^{\prime }=\frac{4k}{
3c^{3}}D^{i\prime \prime }\bar{D}^{j\prime \prime }\delta _{ij},  \label{42}
\end{equation}

\begin{equation}
\psi _{4}^{0}=\sqrt{2}kc^{-5}\left( (D^{i}\bar{D}^{j\prime \prime })^{\prime
\prime \prime }+\frac{2}{3}(D^{i\prime }\bar{D}^{j\prime \prime })^{\prime
\prime }-\frac{1}{3}(\bar{D}^{i\prime \prime }D^{j\prime \prime })^{\prime
}\right) Y_{2ij}^{-2},  \label{43}
\end{equation}

\begin{equation}
\psi _{1}^{0\,k\prime \prime }\ \ =\sqrt{2}kc^{-2}i{\large [}(\bar{D}%
^{i\prime \prime }D^{j\prime \prime })\epsilon _{ijk}+2(D^{i\prime }\bar{D}%
^{j\prime \prime })^{\prime }\epsilon _{ijk}+2iqc\bar{D}^{k\prime \prime
\prime }{\large ]}.  \label{44}
\end{equation}

We have denoted the mass term of the perturbation as $\Upsilon _{\epsilon }$
to indicate that it is a first-order, \textit{perturbative} quantity which
arises as an integrating factor in $\psi _{2}^{0}$. \ The absence of the
Schwarzschild mass is the essential change from the previous section. We now
obtain the reality conditions for this new calculation.

\subsection{Reality Conditions}

As $\psi _{4}^{0}$ has not changed from that obtained in Section 3.1, it
follows that the value of the spin coefficient $\sigma ^{0}$ will remain
that given in Eq.(\ref{23}), so the Bondi mass aspect is simply given as:

\begin{eqnarray}
\Psi &=&\Upsilon _{\epsilon }+{\large (}\frac{2kq}{c^{2}}\bar{D}^{k\prime
\prime }+\frac{\sqrt{2}}{2c}\psi _{1}^{0\,k\prime }+\frac{2ki}{c^{3}}\bar{D}%
^{i\prime \prime }D^{j\prime }\epsilon _{ijk}{\large )}Y_{1k}^{0}  \label{45}
\\
&&+\frac{\sqrt{2}k}{6c^{3}}\dint (D^{i\prime \prime }\bar{D}^{j\prime \prime
})du_{r}Y_{2ij}^{0},  \notag
\end{eqnarray}
and consequently, the reality conditions are little changed from those of
Section 3.2. \ Once again, the $l=2$ condition is trivially satisfied, while
the Bondi mass and linear momentum are given respectively by:

\begin{equation}
\Upsilon _{\epsilon }=\bar{\Upsilon}_{\epsilon }=-\frac{2\sqrt{2}G}{c^{2}}
M_{B}  \label{46}
\end{equation}

\begin{equation}
\Psi ^{k}=-\frac{6G}{c^{3}}P^{k}={\large [}\frac{2kq}{c^{2}}D_{E}^{k\prime }+%
\frac{\sqrt{2}}{2c}\psi _{1R}^{0\,k}-\frac{2k}{c^{3}}(D_{M}^{j\prime
}D_{E}^{i\prime })\epsilon _{ijk}{\large ]}^{\prime }.  \label{47}
\end{equation}
The reality of the $l=1$ coefficient of Eq.(\ref{45}) again yields: 
\begin{equation}
\psi _{1I}^{0\,k\prime }=2\sqrt{2}kqc^{-1}D_{M}^{k\prime \prime }-2\sqrt{2}%
kc^{-2}(D_{E}^{i\prime \prime }D_{E}^{j\prime }+D_{M}^{i\prime \prime
}D_{M}^{j\prime })\epsilon _{ijk}.  \label{psi'}
\end{equation}
\qquad

Aside from Eq.(\ref{46}) these are identical to those obtain in the prior
section.

\subsection{Null Rotations and Equations of Motion}

It is in our attempts to understand the physical content of our equations,
i.e., conservation laws, the definition of angular momentum and spin, and
equations of motion for complex center of mass and charge world-lines that
the present perturbation scheme departs from that of the Schwarzschild
background. \ In particular, using the same stereographic angle field $%
L(u,\zeta ,\bar{\zeta})$ described in Eq.(\ref{LL}) and applying it to the
null rotation of $\psi _{1i}^{0}$ given in Eq.(\ref{34}), we obtain (since $%
M_{B},q$ and $\xi ^{k}$ are first order) the \textit{second order relation}:

\begin{equation}
0=\psi _{1}^{0k}+\frac{6\sqrt{2}G}{c^{2}}M_{B}\xi ^{k}  \label{null.rot2}
\end{equation}
or decomposed as 
\begin{eqnarray}
\psi _{1R}^{0k} &=&-\frac{6\sqrt{2}G}{c^{2}}M_{B}\xi _{R}^{k}  \label{psiR}
\\
\psi _{1I}^{0k} &=&-\frac{6\sqrt{2}G}{c^{2}}M_{B}\xi _{I}^{k}  \label{psiI}
\end{eqnarray}

Taking the prime derivative of Eq.(\ref{psiI}) and inserting it into Eq.(\ref%
{psi'}) we obtain after simplifications, again the classical conservation
law of angular momentum\cite{LL}, 
\begin{eqnarray}
{\large (}M_{B}c\xi _{I}^{k}+\frac{2}{3}c^{-2}qD_{M}^{k\prime }{\large )}%
^{\prime } &=&\frac{2}{3c^{4}}(D_{E}^{i\prime \prime }D_{E}^{j\prime
}+D_{M}^{i\prime \prime }D_{M}^{j\prime })\epsilon _{ijk}  \label{consJ} \\
J^{k\prime } &=&ang.mom.flux  \notag
\end{eqnarray}%
with the identifications 
\begin{eqnarray}
S^{k} &=&M_{B}c\xi _{I}^{k},  \label{S2} \\
J^{k} &=&S^{k}+\frac{2q}{3c}D_{M}^{k\prime }.  \label{J2}
\end{eqnarray}%
This can be considered as the evolution equation for the spin, $S^{k}$.

In order to obtain the dynamical law for the (real) center of mass $\xi
_{R}^{k},$ we first note, from Eq.(\ref{psiR}), that 
\begin{equation*}
M_{B}\xi _{R}^{k\prime \prime }=-\frac{c^{2}}{6\sqrt{2}G}\psi
_{1R}^{0k\prime \prime }.
\end{equation*}
Then, with the use of Eq.(\ref{20e}), i.e.,

\begin{equation}
\psi _{1}^{0\,}{}_{R}^{k\prime \prime }=\sqrt{2}kc^{-2}[2(D_{M}^{j\prime
}D_{E}^{i\prime })^{\prime \prime }-D_{E}^{i\prime \prime }D_{M}^{j\prime
\prime }]\epsilon _{ijk}-2\sqrt{2}kc^{-1}qD_{E}^{k\prime \prime \prime },
\end{equation}%
it becomes 
\begin{equation}
M_{B}\xi _{R}^{k\prime \prime }=-\frac{2}{3c^{4}}[(D_{E}^{i\prime
}D_{M}^{j\prime })^{\prime \prime }-\frac{1}{2}D_{E}^{i\prime \prime
}D_{M}^{j\prime \prime }]\epsilon _{ijk}+\frac{2q}{3c^{-3}}D_{E}^{k\prime
\prime \prime },  \label{motion}
\end{equation}%
a 2$^{nd}$ order differential equation for $\xi _{R}^{k},$ an equation very
much resembling Newton's 2$^{nd}$ law.

\textbf{Remark }Note that if the electric dipole moment had the form $%
D_{E}^{k}=q\xi _{R}^{k},$ i.e., if the center of charge was the same as the
center of mass, the last term would be exactly the classical radiation
reaction force \cite{RadiationReaction}.

Returning to the Bondi linear momentum, Eq.(\ref{47}), after simplification
and the use of Eq.(\ref{psiR}), becomes a dynamical expression for the
momentum:

\begin{equation}
P^{k}=M_{B}\xi _{R}^{k\prime }-\frac{2}{3}c^{-3}qD_{E}^{k\prime \prime }+%
\frac{2}{3}c^{-4}(D_{M}^{j\prime }D_{E}^{i\prime })^{\prime }\epsilon _{ijk}.
\label{P2}
\end{equation}
The dynamical or flux expression for momentum loss is obtained by taking the
prime derivative of Eq.(\ref{P2}) and eliminating the $\xi _{R}^{k\prime
\prime }$ by using Eq.(\ref{motion}), i.e.,

\begin{equation}
P^{k\prime }=\frac{1}{3}c^{-4}D_{E}^{i\prime \prime }D_{M}^{j\prime \prime
}\epsilon _{ijk}.  \label{P'2}
\end{equation}
which is just a different form of Eq.(\ref{motion}).

\subsection{Physical Interpretation}

We thus see that to second order in the perturbation, the imaginary part of
the center of mass world-line leads to exactly the same equation for the
radiated angular momentum as in the Schwarzschild case but, now, with a
first-order (smaller) mass we do obtain recoil. The (real) center of mass
(or the linear momentum) satisfies a 2$^{nd}$ order evolution equation that
is (similar to) Newton's 2$^{nd}$ law with a recoil force and a radiation
reaction force. In some sense we see that general relativity contains
Newton's 2$^{nd}$ law of motion.

In both perturbation schemes there is an anomalous contribution to the total
angular momentum, namely in Eqs.(\ref{J}) and (\ref{J2}) we see the term
proportional to the rate of change of the magnetic dipole moment, i.e., $%
D_{M}^{\prime }.$ This can be considered as a prediction though how to
measure it is not clear.

The two perturbations schemes thus lead to different physical consequences.

\section{Perturbed Metric}

For completeness, we display, to second order, the perturbed
spin-coefficients and metric for the Schwarzschild background calculation.
This metric, given in Bondi coordinates, is obtained by integrating the
Bianchi identities, then integrating the spin coefficient and metric
equations and finally constructing the metric\cite{AdamoNewman}. To begin,
we display the leading-order radial behavior for the spin coefficients; the
more complicated terms are bundled together and given in the Appendix.

\begin{equation}
\kappa =\varepsilon =\pi =0  \label{SC1}
\end{equation}

\begin{equation}
\rho =-\frac{1}{r}-\frac{k\phi _{0}^{0}\bar{\phi}_{0}^{0}}{3r^{5}}
\label{SC2}
\end{equation}

\begin{equation}
\sigma =\frac{\sigma ^{0}}{r^{2}}  \label{SC3}
\end{equation}

\begin{equation}
\tau =-\frac{\psi _{1}^{0}}{2r^{3}}+\mathfrak{S}_{\tau }  \label{SC4}
\end{equation}

\begin{equation}
\alpha =\frac{\alpha ^{0}}{r}-\frac{\beta ^{0}\bar{\sigma}^{0}}{r^{2}}+%
\mathfrak{S}_{\alpha }  \label{SC5}
\end{equation}

\begin{equation}
\beta =\frac{\beta ^{0}}{r}-\frac{\alpha ^{0}\sigma ^{0}}{r^{2}}-\frac{\psi
_{1}^{0}}{2r^{3}}+\mathfrak{S}_{\beta }  \label{SC6}
\end{equation}

\begin{equation}
\gamma =-\frac{\psi _{2}^{0}}{2r^{2}}+\frac{\bar{\eth }\psi _{1}^{0}}{3r^{3}}%
+\mathfrak{S}_{\gamma }  \label{SC7}
\end{equation}

\begin{equation}
\lambda =\frac{\lambda ^{0}}{r}+\frac{\bar{\sigma}^{0}}{r^{2}}+\mathfrak{S}%
_{\lambda }  \label{SC8}
\end{equation}

\begin{equation}
\mu =-\frac{1}{r}-\frac{\psi _{2}^{0}}{r^{2}}+\frac{\bar{\eth }\psi _{1}^{0}%
}{2r^{3}}+\mathfrak{S}_{\mu }  \label{SC9}
\end{equation}

\begin{equation}
\nu =-\frac{\psi _{3}^{0}}{r}+\frac{\bar{\eth }\psi _{2}^{0}}{2r^{2}}-\frac{(%
\bar{\psi}_{1}^{0}+\bar{\eth }^{2}\psi _{1}^{0})}{6r^{3}}+\mathfrak{S}_{\nu }
\label{SC10}
\end{equation}

Non-radial integrating factors in these expressions can be determined from
the relations: 
\begin{eqnarray}
\alpha ^{0} &=&-\bar{\beta}^{0}=\frac{-\zeta }{2},  \label{SC11} \\
\lambda ^{0} &=&\sqrt{2}\bar{\sigma}^{0\prime },  \notag \\
\sigma ^{0} &=&\frac{\sqrt{2}k}{2c^{3}}\left[ \frac{1}{3}\dint (D^{i\prime
\prime }\bar{D}^{j\prime \prime })du_{r}-(\bar{D}^{i}D^{j\prime \prime
})^{\prime }-\frac{2}{3}(\bar{D}^{i\prime }D^{j\prime \prime })\right]
Y_{2ij}^{2}.  \notag \\
\psi _{3}^{0} &=&\sqrt{2}\eth \bar{\sigma}^{0\prime },  \notag \\
\psi _{4}^{0} &=&-\sqrt{2}\lambda ^{0\prime }=-2\bar{\sigma}^{0\prime \prime
}  \notag
\end{eqnarray}

It is useful to define an auxiliary bundled expression, $\mathfrak{A}$,
which enters into the metric expressions: 
\begin{eqnarray}
\mathfrak{A~} &\mathfrak{=}&\mathfrak{~}\frac{\bar{\eth }\psi _{1}^{0}+\eth 
\bar{\psi}_{1}^{0}}{6r^{2}}-\frac{kq^{2}}{r^{2}}-kq\left( \frac{\sqrt{2}%
(D^{i\prime }+\bar{D}^{i\prime })}{r^{2}}+\frac{D^{i}-\bar{D}^{i}}{3r^{3}}%
\right) Y_{1i}^{0}  \label{M1} \\
&&-k\left( \frac{10D^{i\prime }\bar{D}^{j\prime }}{9r^{2}}-\frac{(1+7\sqrt{2}%
)(D^{i}\bar{D}^{j})^{\prime }}{27r^{3}}+\frac{13D^{i}\bar{D}^{j}}{45r^{4}}-%
\frac{8\sqrt{2}GM_{S}D^{i}\bar{D}^{j}}{45c^{2}r^{5}}\right) \delta _{ij} 
\notag \\
&&-ik\left( \frac{2D^{i}\bar{D}^{j\prime }-(D^{i}\bar{D}^{j})^{\prime }}{%
3r^{3}}\right) \epsilon _{ijk}Y_{1k}^{0}-k\left( \frac{2D^{i\prime }\bar{D}%
^{j\prime }}{3r^{2}}+\frac{5\sqrt{2}D^{i}\bar{D}^{j\prime }}{18r^{3}}\right)
Y_{2ij}^{0}  \notag \\
&&-k\left( \frac{(1-4\sqrt{2})(D^{i}\bar{D}^{j})^{\prime }}{108r^{3}}-\frac{%
7D^{i}\bar{D}^{j}}{90r^{4}}+\frac{2\sqrt{2}GM_{S}D^{i}\bar{D}^{j}}{%
45c^{2}r^{5}}\right) Y_{2ij}^{0}.  \notag
\end{eqnarray}

\qquad The full covariant metric to second order in the Schwarzschild
perturbation, takes the form

$\qquad \qquad \qquad \qquad \qquad \qquad g_{ab}=\left[ 
\begin{array}{cccc}
g_{00} & 1 & g_{0\zeta } & g_{0\overline{\zeta }} \\ 
1 & 0 & 0 & 0 \\ 
g_{0\zeta } & 0 & g_{\zeta \zeta } & g_{\zeta \overline{\zeta }} \\ 
g_{0\overline{\zeta }} & 0 & g_{\zeta \overline{\zeta }} & g_{\overline{%
\zeta }\overline{\zeta }}%
\end{array}%
\right] $

with: 
\begin{eqnarray}
g_{00} &=&2+\frac{2}{r}{\large (}\Upsilon _{\epsilon }-\frac{2\sqrt{2}GM_{S}%
}{c^{2}}{\large )}+\frac{\sqrt{2}}{2r}(\psi _{1}^{0k\prime }+\bar{\psi}%
_{1}^{0k\prime })Y_{1k}^{0}  \label{M2} \\
&&-\frac{2ik}{r}(D^{i\prime }\bar{D}^{j\prime })^{\prime }\epsilon
_{ijk}Y_{1k}^{0}-2\mathfrak{A}  \notag \\
&&+\frac{k}{r}{\large (}\frac{\sqrt{2}}{2}(D^{i}\bar{D}^{j\prime \prime }+%
\bar{D}^{i}D^{j\prime \prime })^{\prime }+\frac{\sqrt{2}}{3}(D^{i\prime }%
\bar{D}^{j\prime })^{\prime }{\large )}Y_{2ij}^{0},  \notag
\end{eqnarray}

\begin{eqnarray}
g_{\zeta 0} &=&P^{-1}{\large [}\bar{\eth }\sigma ^{0}+\frac{2\psi
_{1}^{0k}Y_{1k}^{1}}{3r}-\frac{2kqD^{i}Y_{1i}^{1}}{r^{2}}{\large ]}
\label{M3} \\
&&-\frac{ik}{P}{\large (}\frac{2D^{i}\bar{D}^{j\prime }}{r^{2}}-\frac{4\sqrt{%
2}D^{i}\bar{D}^{j}}{15r^{3}}{\large )}\epsilon _{ijk}Y_{1k}^{1}  \notag \\
&&+\frac{k}{P}{\large (}\frac{2D^{i}\bar{D}^{j\prime \prime }}{r}-\frac{4%
\sqrt{2}D^{i}\bar{D}^{j\prime }}{3r^{2}}+\frac{4D^{i}\bar{D}^{j}}{15r^{3}}%
{\large )}Y_{2ij}^{1},  \notag
\end{eqnarray}

\begin{eqnarray}
g_{\bar{\zeta}0} &=&P^{-1}{\large [}\eth \bar{\sigma}^{0}+\frac{2\bar{\psi}%
_{1}^{0k}Y_{1k}^{-1}}{3r}-\frac{2kq\bar{D}^{i}Y_{1i}^{-1}}{r^{2}}{\large ]}
\label{M4} \\
&&+\frac{ik}{P}{\large (}\frac{2\bar{D}^{i}D^{j\prime }}{r^{2}}-\frac{4\sqrt{%
2}\bar{D}^{i}D^{j}}{15r^{3}}{\large )}\epsilon _{ijk}Y_{1k}^{-1}  \notag \\
&&+\frac{k}{P}{\large (}\frac{2\bar{D}^{i}D^{j\prime \prime }}{r}-\frac{4%
\sqrt{2}\bar{D}^{i}D^{j\prime }}{3r^{2}}+\frac{4\bar{D}^{i}D^{j}}{15r^{3}}%
{\large )}Y_{2ij}^{-1},  \notag
\end{eqnarray}

\begin{equation}
g_{\zeta \zeta }=-\frac{2\sigma ^{0}r}{P^{2}},~~~~~g_{\bar{\zeta}\bar{\zeta}%
}=-\frac{2\bar{\sigma}^{0}r}{P^{2}},  \label{M5}
\end{equation}

\begin{equation}
g_{\zeta \bar{\zeta}}=\frac{-r^{2}}{P^{2}}.  \label{M7}
\end{equation}%
and%
\begin{equation*}
P\equiv (1+\zeta \bar{\zeta}).
\end{equation*}

The corresponding metric for the perturbation off of the Minkowski
background is easily obtained from these results.

\section{Discussion}

We begin with a mea culpa. In this work we started with two given vacuum
metrics, the Schwarzschild and Minkowski metrics and then `drove' or
perturbed them both with a `small' Maxwell field. \ In one case there was a
`large' mass while in the second case the mass was `small'. \ There was no
systematic attempt, by comparisons, to define small or large quantities;
there was no small parameter for a series expansion. \ By 'small' we meant
it to be as small as needed for a physical effect; or as large as was needed
for a different effect. The idea behind the two perturbation schemes was
purely heuristic. \ We wanted to see, in a rough sense, within the \textit{%
context of the Einstein equations\cite{PhysicalContent},} what would be the
physical responses of the recently defined (complex) \textit{center of mass}
of massive systems (with different size masses) to an electromagnetic
perturbation. In addition to the totally expected result that the large mass
object did not experience any recoil while the small object did experience
recoil, we obtained a variety of concomitant physical results. The imaginary
part of the (complex) \textit{center of mass world-line }could be identified
with the electrodynamic induced internal spin angular momentum that
satisfies the classical (angular momentum) conservation law. There was an
electromagnetically induced gravitational radiation as well as linear
momentum loss. \ We obtained explicit equations of motion for the center of
mass that have the form of Newton's 2$^{nd}$ law.

The results presented here, in the two different models, are in total
agreement with well-known classical Newtonian and electrodynamics effects. \
They however go beyond these known results. \ They give a confirmation of
the physical ideas developed purely in the context gravitational theory
(general relativity) for the identification and the associated dynamical
response of certain geometric structures that arise naturally. \ The most
important of them is the special class of null geodesic congruences, the
shear-free (or asymptotically shear-free) null geodesic congruences. \ It is
their very existence that yields or provides the complex Minkowski
world-lines that become our complex center-of mass which then yield our
angular momentum expressions.

As a final comment we point out that the perturbed space-times described
here contain, as special cases, both the Kerr and the charged Kerr metrics
as perturbations off Schwarzschild. They occur when the electromagnetic
dipoles are `shut off', i.e., when $D^{i}=0.$

\section{Acknowledgments}

We take this opportunity to thank Carlos Kozameh and Gilberto Silva-Ortigoza
for the many long and detailed discussions that clarified many of the ideas
that preceded the present work.

\section{Appendix}

\subsection{Radial Behavior of the Weyl Tensor}

Below are the full expressions for the bundle terms ($\mathcal{A}_{i}$)
given in Eq. \ref{8}-\ref{11}: 
\begin{eqnarray}
\mathcal{A}_{1} &=&-\frac{5k}{r^{5}}{\Large [}2qD^{i}Y_{1i}^{1}+2\sqrt{2}%
D^{i}\bar{D}^{j\prime }{\large (}\frac{i}{\sqrt{2}}\epsilon _{ijk}Y_{1k}^{1}+%
\frac{1}{2}Y_{2ij}^{1}{\large )}{\Large ]}  \label{A1} \\
&&+\frac{6kD^{i}\bar{D}^{j}}{r^{6}}{\large (}\frac{i}{\sqrt{2}}\epsilon
_{ijk}Y_{1k}^{1}+\frac{1}{2}Y_{2ij}^{1}{\large )}  \notag
\end{eqnarray}

\begin{eqnarray}
\mathcal{A}_{2} &=&-\frac{\bar{\eth }\psi _{1}^{0}}{r^{4}}+\frac{2kq^{2}}{%
r^{4}}+kq{\large (}\frac{2\sqrt{2}(D^{i\prime }+\bar{D}^{i\prime })}{r^{4}}+%
\frac{10D^{i}-2\bar{D}^{i}}{3r^{5}}{\large )}Y_{1i}^{0}  \label{A2} \\
&&+2k{\large (}\frac{D^{i\prime }\bar{D}^{j\prime }}{r^{4}}-\frac{(1+4\sqrt{2%
})(D^{i}\bar{D}^{j})^{\prime }}{9r^{5}}+\frac{10D^{i}\bar{D}^{j}}{9r^{6}}-%
\frac{8\sqrt{2}GM_{S}D^{i}\bar{D}^{j}}{3c^{2}r^{7}}{\large )}\delta _{ij} 
\notag \\
&&+ik{\large (}\frac{12D^{i}\bar{D}^{j\prime }}{3r^{5}}+\frac{(4+\sqrt{2}%
)(D^{i}\bar{D}^{j})^{\prime }}{6r^{5}}-\frac{\sqrt{2}D^{i}\bar{D}^{j}}{r^{6}}%
{\large )}\epsilon _{ijk}Y_{1k}^{0}  \notag \\
&&+k{\large (}\frac{4D^{i\prime }\bar{D}^{j\prime }}{3r^{4}}+\frac{30\sqrt{2}%
D^{i}\bar{D}^{j\prime }+(1+2\sqrt{2})(D^{i}\bar{D}^{j})^{\prime }}{18r^{5}}%
{\large )}Y_{2ij}^{0}  \notag \\
&&-k{\large (}\frac{10D^{i}\bar{D}^{j}}{9r^{6}}-\frac{2\sqrt{2}GM_{S}D^{i}%
\bar{D}^{j}}{3c^{2}r^{7}}{\large )}Y_{2ij}^{0}  \notag
\end{eqnarray}

\begin{eqnarray}
\mathcal{A}_{3} &=&-\frac{\bar{\eth }\psi _{2}^{0}}{r^{3}}+\frac{\bar{\eth }%
^{2}\psi _{1}^{0}}{2r^{4}}-\frac{2kqD^{i\prime \prime }}{r^{3}}Y_{1i}^{-1}
\label{A3} \\
&&+2kq{\large (}\frac{(1+7\sqrt{2})D^{i\prime }+6\sqrt{2}\bar{D}^{i\prime }}{%
3r^{4}}+\frac{(11D^{i}+12\bar{D}^{i})}{18r^{5}}-\frac{2\sqrt{2}GM_{S}\bar{D}%
^{i}}{c^{2}r^{6}}{\large )}Y_{1i}^{-1}  \notag \\
&&+ik{\large (}\frac{2D^{i\prime \prime }\bar{D}^{j\prime }}{r^{3}}-\frac{2%
\sqrt{2}D^{i\prime }\bar{D}^{j\prime }+(2-2\sqrt{2})(\bar{D}^{i}D^{j\prime
})^{\prime }}{3r^{4}}{\large )}\epsilon _{ijk}Y_{1k}^{-1}  \notag \\
&&+ik{\large (}\frac{(34-\sqrt{2})(D^{i}\bar{D}^{j})^{\prime }-D^{i}\bar{D}%
^{j\prime }}{9r^{5}}+\frac{5\sqrt{2}\bar{D}^{i}D^{j}}{4r^{6}}-\frac{56GM_{S}%
\bar{D}^{i}D^{j}}{5c^{2}r^{7}}{\large )}\epsilon _{ijk}Y_{1k}^{-1}  \notag \\
&&-k{\large (}\frac{\sqrt{2}D^{i\prime \prime }\bar{D}^{j\prime }}{r^{3}}-%
\frac{20D^{i\prime }\bar{D}^{j\prime }+\sqrt{2}(\bar{D}^{i}D^{j\prime
})^{\prime }}{3r^{4}}{\large )}Y_{2ij}^{-1}  \notag \\
&&-k{\large (}\frac{37\sqrt{2}D^{i\prime }\bar{D}^{j}+(3-49\sqrt{2})(D^{i}%
\bar{D}^{j})^{\prime }}{18r^{5}}+\frac{23D^{i}\bar{D}^{j}}{12r^{6}}{\large )}%
Y_{2ij}^{-1}  \notag \\
&&+\frac{2\sqrt{2}Gk}{c^{2}}{\large (}\frac{\sqrt{2}M_{S}\bar{D}%
^{i}D^{j\prime }}{r^{6}}+\frac{16M_{S}D^{i}\bar{D}^{j}}{5r^{7}}{\large )}%
Y_{2ij}^{-1}  \notag
\end{eqnarray}

\begin{eqnarray}
\mathcal{A}_{4} &=&-\frac{\bar{\eth }\psi _{3}^{0}}{r^{2}}+\frac{\bar{\eth }%
^{2}\psi _{2}^{0}}{2r^{3}}-\frac{\bar{\eth }^{3}\psi _{1}^{0}}{6r^{4}}
\label{A4} \\
&&-k{\Huge (}\frac{3\sqrt{2}D^{i\prime \prime }\bar{D}^{j\prime
}+2(D^{i\prime \prime }\bar{D}^{j})^{\prime }}{r^{3}}-\frac{(16\sqrt{2}
-72)(D^{i\prime }\bar{D}^{j})^{\prime }}{9r^{4}}  \notag \\
&&-\frac{128D^{i\prime }\bar{D}^{j\prime }}{9r^{4}}+\frac{(24-107\sqrt{2}
)(D^{i}\bar{D}^{j})^{\prime }-25\sqrt{2}D^{i\prime }\bar{D}^{j}}{36r^{5}}+%
\frac{47D^{i}\bar{D}^{j}}{15r^{6}}  \notag \\
&&+\frac{240GM_{S}D^{i\prime }\bar{D}^{j}}{15c^{2}r^{6}}-\frac{38\sqrt{2}
GM_{S}D^{i}\bar{D}^{j}}{5c^{2}r^{7}}{\Huge )}Y_{2ij}^{-2}.  \notag
\end{eqnarray}

\subsection{The Spin Coefficients}

The following are the full expressions for the $\mathfrak{S}_{i}$ given in
Section V for the spin coefficients in the Schwarzschild background
perturbation:

\begin{eqnarray}
\mathfrak{S}_{\tau } &=&\frac{8kqD^{i}}{3r^{4}}Y_{1i}^{1}+ik{\large (}\frac{%
8D^{i}\bar{D}^{j\prime }}{3r^{4}}-\frac{\sqrt{2}D^{i}\bar{D}^{j}}{2r^{5}}%
{\large )}\epsilon _{ijk}Y_{1k}^{1}  \label{ST} \\
&&+k{\large (}\frac{4\sqrt{2}D^{i}\bar{D}^{j\prime }}{3r^{4}}-\frac{D^{i}%
\bar{D}^{j}}{2r^{5}}{\large )}Y_{2ij}^{1}  \notag
\end{eqnarray}

\begin{eqnarray}
\mathfrak{S}_{\alpha } &=&-\frac{2kq\bar{D}^{i}}{3r^{4}}Y_{1i}^{-1}+ik%
{\large (}\frac{2\bar{D}^{i}D^{j\prime }}{3r^{4}}-\frac{\sqrt{2}\bar{D}%
^{i}D^{j}}{4r^{5}}{\large )}\epsilon _{ijk}Y_{1k}^{-1}  \label{SA} \\
&&-k{\large (}\frac{\sqrt{2}\bar{D}^{i}D^{j\prime }}{3r^{4}}-\frac{\bar{D}%
^{i}D^{j}}{4r^{5}}{\large )}Y_{2ij}^{-1}+\frac{k\alpha ^{0}\phi _{0}^{0}\bar{%
\phi}_{0}^{0}}{12r^{5}}  \notag
\end{eqnarray}

\begin{eqnarray}
\mathfrak{S}_{\beta } &=&\frac{10kqD^{i}}{3r^{4}}Y_{1i}^{1}+ik{\large (}%
\frac{10D^{i}\bar{D}^{j\prime }}{3r^{4}}-\frac{\sqrt{2}D^{i}\bar{D}^{j}}{%
2r^{5}}{\large )}\epsilon _{ijk}Y_{1k}^{1}  \label{SB} \\
&&+k{\large (}\frac{5\sqrt{2}D^{i}\bar{D}^{j\prime }}{3r^{4}}-\frac{D^{i}%
\bar{D}^{j}}{2r^{5}}{\large )}Y_{2ij}^{1}+\frac{k\beta ^{0}\phi _{0}^{0}\bar{%
\phi}_{0}^{0}}{12r^{5}}  \notag
\end{eqnarray}

\begin{eqnarray}
\mathfrak{S}_{\gamma } &=&\frac{\alpha ^{0}\psi _{1}^{0}+\beta ^{0}\bar{\psi}%
_{1}^{0}}{6r^{3}}-\frac{kq^{2}}{r^{3}}-kq{\large (}\frac{\sqrt{2}(D^{i\prime
}+\bar{D}^{i\prime })}{r^{3}}+\frac{7D^{i}-5\bar{D}^{i}}{12r^{4}}{\large )}%
Y_{1i}^{0}  \label{SG} \\
&&-k{\large (}\frac{10D^{i\prime }\bar{D}^{j\prime }}{9r^{3}}-\frac{(1+7%
\sqrt{2})(D^{i}\bar{D}^{j})^{\prime }}{18r^{4}}+\frac{26D^{i}\bar{D}^{j}}{%
45r^{5}}+\frac{2MD^{i}\bar{D}^{j}}{9r^{6}}{\large )}\delta _{ij}  \notag \\
&&-ik{\large (}\frac{24D^{i}\bar{D}^{j\prime }+(4+\sqrt{2})(D^{i}\bar{D}%
^{j})^{\prime }}{24r^{4}}-\frac{\sqrt{2}D^{i}\bar{D}^{j}}{6r^{6}}{\large )}%
\epsilon _{ijk}Y_{1k}^{0}  \notag \\
&&-k{\large (}\frac{2D^{i\prime }\bar{D}^{j\prime }}{3r^{3}}+\frac{(1-4\sqrt{%
2})(D^{i}\bar{D}^{j})^{\prime }+30\sqrt{2}D^{i}\bar{D}^{j\prime }}{24r^{4}}-%
\frac{7D^{i}\bar{D}^{j}}{45r^{5}}-\frac{MD^{i}\bar{D}^{j}}{18r^{6}}{\large )}%
Y_{2ij}^{0}+\mathfrak{B}  \notag
\end{eqnarray}

\begin{eqnarray}
\mathfrak{B} &\equiv &-\frac{2kq}{3r^{4}}(\alpha ^{0}D^{i}Y_{1i}^{1}+\beta
^{0}\bar{D}^{i}Y_{1i}^{-1})-ik\alpha ^{0}{\large (}\frac{8D^{i}\bar{D}%
^{j\prime }}{9r^{3}}-\frac{\sqrt{2}D^{i}\bar{D}^{j}}{16r^{4}}{\large )}%
\epsilon _{ijk}Y_{1k}^{1}  \label{B} \\
&&-k\alpha ^{0}{\large (}\frac{4\sqrt{2}D^{i}\bar{D}^{j\prime }}{9r^{3}}-%
\frac{D^{i}\bar{D}^{j}}{16r^{4}}{\large )}Y_{2ij}^{1}+ik\beta ^{0}{\large (}%
\frac{8\bar{D}^{i}D^{j\prime }}{9r^{3}}-\frac{\sqrt{2}\bar{D}^{i}D^{j}}{%
16r^{4}}{\large )}\epsilon _{ijk}Y_{1k}^{-1}  \notag \\
&&-k\beta ^{0}{\large (}\frac{4\sqrt{2}\bar{D}^{i}D^{j\prime }}{9r^{3}}-%
\frac{\bar{D}^{i}D^{j}}{16r^{5}}{\large )}Y_{2ij}^{-1}  \notag
\end{eqnarray}

\begin{equation}
\mathfrak{S}_{\lambda }=k{\large (}\frac{2D^{i\prime \prime }\bar{D}^{j}}{%
r^{3}}-\frac{4\sqrt{2}D^{i\prime }\bar{D}^{j}}{3r^{4}}+\frac{D^{i}\bar{D}^{j}%
}{2r^{5}}{\large )}Y_{2ij}^{-2}  \label{SL}
\end{equation}

\begin{eqnarray}
\mathfrak{S}_{\mu } &=&-\frac{kq^{2}}{r^{3}}-kq{\large (}\frac{\sqrt{2}%
(D^{i\prime }+\bar{D}^{i\prime })}{r^{3}}+\frac{10D^{i}-2\bar{D}^{i}}{9r^{4}}%
{\large )}Y_{1i}^{0}  \label{SM} \\
&&-k{\large (}\frac{D^{i\prime }\bar{D}^{j\prime }}{r^{3}}-\frac{(2+8\sqrt{2}%
)(D^{i}\bar{D}^{j})^{\prime }}{27r^{4}}+\frac{5D^{i}\bar{D}^{j}}{3r^{5}}-%
\frac{8\sqrt{2}GM_{S}D^{i}\bar{D}^{j}}{15c^{2}r^{6}}{\large )}\delta _{ij} 
\notag \\
&&-ik{\large (}\frac{4D^{i}\bar{D}^{j\prime }}{3r^{4}}+\frac{(4+\sqrt{2}%
)(D^{i}\bar{D}^{j})^{\prime }}{18r^{4}}-\frac{\sqrt{2}D^{i}\bar{D}^{j}}{%
4r^{5}}{\large )}\epsilon _{ijk}Y_{1k}^{0}-\frac{2\sqrt{2}kGM_{S}D^{i}\bar{D}%
^{j}}{15c^{2}r^{6}}Y_{2ij}^{0}  \notag \\
&&-k{\large (}\frac{2D^{i\prime }\bar{D}^{j\prime }}{3r^{3}}+\frac{(1+2\sqrt{%
2})(D^{i}\bar{D}^{j})^{\prime }+30\sqrt{2}D^{i}\bar{D}^{j\prime }}{54r^{4}}-%
\frac{10D^{i}\bar{D}^{j}}{36r^{5}}{\large )}Y_{2ij}^{0}  \notag
\end{eqnarray}

\begin{eqnarray}
\mathfrak{S}_{\nu } &=&-ik{\large (}\frac{2D^{i\prime \prime }\bar{D}%
^{j\prime }}{r^{2}}-\frac{8\sqrt{2}D^{i\prime }\bar{D}^{j\prime }+\sqrt{2}%
D^{i\prime \prime }\bar{D}^{j}+(1-\sqrt{2})(\bar{D}^{i}D^{j\prime })^{\prime
}}{9r^{3}}{\large )}\epsilon _{ijk}Y_{1k}^{-1}  \label{SN} \\
&&-ik{\large (}\frac{8D^{i}\bar{D}^{j\prime }+6\bar{D}^{i}D^{j\prime }+(34-%
\sqrt{2})(D^{i}\bar{D}^{j})^{\prime }}{36r^{4}}-\frac{7\sqrt{2}\bar{D}%
^{i}D^{j}}{20r^{5}}{\large )}\epsilon _{ijk}Y_{1k}^{-1}  \notag \\
&&-ik{\large (}\frac{8M\bar{D}^{i}D^{j\prime }}{15r^{5}}-\frac{92GM_{S}\bar{D%
}^{i}D^{j}}{60c^{2}r^{6}}{\large )}\epsilon _{ijk}Y_{1k}^{-1}+k{\large (}%
\frac{\sqrt{2}D^{i\prime \prime }\bar{D}^{j\prime }}{r^{2}}{\large )}%
Y_{2ij}^{-1}  \notag \\
&&-k{\large (}\frac{3D^{i\prime \prime }\bar{D}^{j}+26D^{i\prime }\bar{D}%
^{j\prime }+\sqrt{2}(\bar{D}^{i}D^{j\prime })^{\prime }}{9r^{3}}-\frac{70%
\sqrt{2}D^{i\prime }\bar{D}^{j}+(3-40\sqrt{2})(D^{i}\bar{D}^{j})^{\prime }}{%
72r^{4}}{\large )}Y_{2ij}^{-1}  \notag \\
&&+k{\large (}\frac{11c^{2}\bar{D}^{i}D^{j}-16GM_{S}\bar{D}^{i}D^{j\prime }}{%
60c^{2}r^{5}}-\frac{54\sqrt{2}GM_{S}D^{i}\bar{D}^{j}}{60c^{2}r^{6}}{\large )}%
Y_{2ij}^{-1}.  \notag
\end{eqnarray}

\end{document}